\documentclass[12pt]{iopart}
\usepackage{braket}
\usepackage{bm}
\usepackage{graphicx}
\usepackage{iopams}
\newcommand{\bvec}[1]{{\boldsymbol{#1}}}

\begin{document}

\title{Spin correlations and entanglement in partially magnetised ensembles of fermions}
\author{G. S. Thekkadath$^1$, Liang Jiang$^2$, J. H. Thywissen$^{1,3}$}
\address{$^1$ Department of Physics and CQIQC, University of Toronto, M5S 1A7 Canada}
\address{$^2$ Department of Applied Physics, Yale University, New Haven, CT 06511 USA}
\address{$^3$ Canadian Institute for Advanced Research, Toronto, Ontario M5G 1Z8 Canada}
\date{\today}

\begin{abstract}
We show that the singlet fraction $p_s$ and total magnetisation (or polarisation) $m$ can bound the minimum concurrence in an ensemble of spins. We identify $p_s > (1-m^2)/2$ as a sufficient and tight condition for bipartite entanglement. Our proof makes no assumptions about the state of the system or symmetry of the particles, and can therefore be used as a witness for spin entanglement between fermions. We discuss the implications for recent experiments in which spin correlations were observed, and the prospect to study entanglement dynamics in the demagnetisation of a cold Fermi gas.
\end{abstract}

\maketitle

\section{Introduction}

Spin correlations have recently been observed in cold fermionic atoms as signatures of pairing, magnetism, and interaction strength \cite{greif_short-range_2013,bardon_transverse_2014,hart_observation_2015}. It is interesting to ask if the observed correlations require pairwise entanglement. Since typical experimental samples contain thousands of atoms, full tomography is inaccessible; instead, one must find an entanglement witness based on a reduced set of measurements \cite{Lewenstein:2007hr,guhne_entanglement_2009}. A commonly explored approach has been to measure spin squeezing \cite{Sorensen:2001jn,wang2003spin,korbicz_spin_2005,kiesel_experimental_2007,Toth:2007ix,Esteve:2008ij,baragiola2010collective}, however this approach is mainly limited to symmetric states or indistinguishable particles \cite{Sorensen:2001jn,wang2003spin,korbicz_spin_2005}, and thus inapplicable to spin mixtures of fermions.

An alternate characterisation may come from the degree of polarisation (or magnetisation) $m$, and the spin-singlet fraction $p_s$ of the ensemble (see Sec.~\ref{sec:definitions} for precise definitions). The singlet state plays a key role in the physics of ultracold fermions, since a singlet spin wave function is required for s-wave interactions, which are the only interactions not suppressed at low energy by the centrifugal barrier. 
For spin mixtures near a Feshbach resonance \cite{Chin2010}, the pairing fraction can be measured by an adiabatic rapid passage that projects interacting pairs onto molecular dimers \cite{strecker_conversion_2003,pazy_conversion_2004,hodby_production_2005}. 
This enables direct measurement of $p_s$ in an ensemble. Singlet fraction is also proportional to the s-wave contact \cite{Partridge:2005jh,Tan:2008ey,Tan:2008eg,Tan:2008ch,Braaten:2008tc,Zhang:2009kq,Werner:2009vb,Stewart:2010fy,Vale2010,bardon_transverse_2014},
and singlet pairs in an optical superlattice can also be mapped or projected onto excited motional states \cite{Rey:2007gc,anderlini_controlled_2007,Trotzky:2008jy,Trotzky:2010ka,Greif:2011fr,Nascimbene:2012fp,greif_short-range_2013}. 

It is well known that $p_s > 1/2$ indicates pairwise entanglement in unpolarised ($m=0$) ensembles \cite{Werner:1989hn}. The existence of a threshold is intuitive, since spin singlets are antisymmetric Bell states. 
Here we assume both $m$ and $p_s$ of an ensemble are measured, but make no assumptions about the form of the reduced two-body density operator $\hat{\rho}^{AB}$, which has 15 degrees of freedom.

We find that the concurrence $\mathcal{C}$ of the ensemble can be bounded:
\begin{equation}
\label{eq:minconc}
\mathcal{C} \geq \mathrm{max}\left[p_s - \sqrt{(1-p_s)^2 - m^2},0 \right].
\end{equation}
This delineates a bound on the singlet fraction of an arbitrary two-body state that is a {\em sufficient and tight condition for its entanglement}, namely
\begin{equation} \label{eq:singlet_bound}
p_s > \frac{1-m^2}{2}.
\end{equation}
This ``singlet bound'' is an extension of the Werner bound to partially polarised ($m>0$) ensembles, where $p_s>1/2$ is sufficient but is not a tight bound. Our proof makes no assumptions about the state and therefore is a condition for bipartite entanglement in any ensemble of spins. Thus, Eq.~(\ref{eq:minconc}) can elucidate the nature of spin correlations in recent experiments with interacting ensembles of spin-half fermions, even in states far from equilibrium.

\section{One- and two-body observables \label{sec:definitions}}

The spin state of pairs in a spin-1/2 ensemble can be described using the antisymmetric singlet state $\ket{s_0}=(\ket{\uparrow \downarrow} - \ket{\downarrow \uparrow})/\sqrt{2}$ and the symmetric triplet states $\ket{t_0}=(\ket{\uparrow \downarrow}+\ket{\downarrow \uparrow})/\sqrt{2}$, $\ket{t_1} = \ket{\uparrow \uparrow}$, $\ket{t_{-1}} = \ket{\downarrow \downarrow}$. These form an orthonormal set of basis states with well defined angular momentum quantum numbers $\ket{S,S_z}$. The most general state can be written 
\begin{eqnarray}
\label{eq:general_matrix}
\fl \qquad \qquad \hat{\rho}^{AB} = p_s \ket{s_0}\bra{s_0} + \sum_{i \in \left \{ 0, \pm 1 \right \}} (q_i\ket{s_0}\bra{t_i} + q_i^*\ket{t_i}\bra{s_0}) + \sum_{i,j \in \left \{ 0, \pm 1 \right \}} (p_{ij}\ket{t_i}\bra{t_j})
\end{eqnarray}
where the populations are normalised to $Tr[\hat{\rho}^{AB}]=1$, and $p_s$ is the singlet fraction.

The magnetisation is $\bvec{m} = (m_x, m_y, m_z) = Tr [\bvec{\hat{S}}\hat{\rho}^{AB}]$, where $\bvec{\hat{S}} = (\bvec{\hat{\sigma}}^A + \bvec{\hat{\sigma}}^B)/2$ and $\bvec{\hat{\sigma}}^{A,B}$ are the usual Pauli spin operators. The reduced one-body states (e.g., $\hat{\rho}_A = Tr_B \hat{\rho} ^{AB}$) are completely defined by a Bloch vector $\bvec{v}$: $\hat{\rho}_{A,B} \to \hat{I}/2 + \bvec{v}_{A,B} \cdot \bvec{\hat{\sigma}}/2$, in which 
$\hat{I}$ is the identity operator. Since $\bvec{m} = \case{1}{2} \bvec{v}_A + \case{1}{2}\bvec{v}_B$,
\begin{equation}
\label{eq:bloch_mag_mixed}
m^2 = \case{1}{4}(v_A^2 + v_B^2 + 2v_A  v_B\cos{\beta})
\end{equation}
where $\beta$ is the angle between the two Bloch vectors, and $m=|\bvec{m}|$.

In the problem we are considering, only the ensemble observables $\bvec{m}$ and $p_s$ are measured, not $v_A$, $v_B$, or $\beta$. One simple relation between $m$ and $p_s$ is given by the normalisation of probability:
\begin{equation}
\label{eq:physical_bound}
p_s \leq 1 - m.
\end{equation}
This can be shown by noting that $m_z =  p_{11} - p_{-1-1}$ and $p_s + \sum_i p_{ii} = 1$, from which the singlet fraction is bounded by $p_s \leq 1 - |m_z| - p_{00}$. Since $|m_z| \leq m$, Eq.~(\ref{eq:physical_bound}) follows. 

\section{Unentangled spins \label{sec:unentangled}}

Let us start by finding the singlet fraction of the separable state $\hat{\rho}^{AB} = \hat{\rho}^A \otimes \hat{\rho}^B$ where $\hat{\rho}^A$ and $\hat{\rho}^B$ can be different mixed states. Since we are seeking a relation between two rotationally invariant quantities, $m$ and $p_s$, we are free to choose the coordinate system, and align $\hat{\rho}^A$ along the $z$ axis in Bloch space. Then $\hat{\rho}^A = p_{A \uparrow} \ket{\uparrow_A} \bra{\uparrow_A} + p_{A \downarrow} \ket{\downarrow_A}\bra{\downarrow_A}$, whereas $\hat{\rho}^B$ remains arbitrary, and we write it as $\hat{\rho}^B = \sum_{ij} c_{ij}\ket{i_B}\bra{j_B}$ where $i,j \in \left \{ \uparrow,\downarrow \right \}$. The singlet fraction is
\begin{equation}
p_s  = \case{1}{2} p_{A \uparrow}c_{\downarrow\downarrow} + \case{1}{2}  p_{A \downarrow}c_{\uparrow \uparrow}.
\end{equation}
In terms of the Bloch vectors, $p_{A \uparrow} = (1+v_A)/2$ and $p_{A \downarrow} = (1-v_A)/2$, whereas $c_{\downarrow\downarrow} = (1-v_B \cos{\beta})/2$ and $c_{\uparrow \uparrow} = (1+v_B \cos{\beta})/2$, thus
\begin{equation} \label{eq:ps}
p_s =  \case{1}{4}(1-v_A v_B\cos{\beta}).
\end{equation}
Equation~(\ref{eq:ps}) has a simple interpretation for two pure states: when the first spin is along the $+z$ axis of the Bloch sphere, the antiparallel (spin-down) fraction of the second spin is equally split between singlets and triplet zeros \cite{AMSpc}.

For an ensemble of unentangled qubits, each of which is in the same unknown mixed state, Gisin noted that $p_s = (1-m^2)/4$, and proposed measuring $p_s$ as a more efficient determination of $m$ than measuring $\bvec{m}$   \cite{gisin}. We recover this result from Eq.~(\ref{eq:ps}) with $v_A = v_B$ and $\beta=0$. However for arbitrary $\bvec{v}_A$ or $\bvec{v}_B$, we can only bound $p_s$: eliminating $\beta$, with Eq.~(\ref{eq:bloch_mag_mixed}),
\begin{equation}
\label{eq:vectorial_result}
p_s = \case{1}{2} \big( 1-m^2 +\case{1}{4}(v_A^2 - 1 + v_B^2 - 1 ) \big ) \leq \frac{1-m^2}{2},
\end{equation}
where the inequality holds because $|v_A|\leq1$ and $|v_B|\leq1$. Note that separable pure ($v_A = v_B = 1$) states are examples of non-entangled states on the line $p_s = (1-m^2)/2$, which demonstrates the tightness of Eq.~(\ref{eq:singlet_bound}). (If however magnetisation is known only along $z$, but the full magnetisation possibly lies along another direction, the singlet bound $p_s > (1 - m_z^2)/2$ is sufficient but no longer tight.)

We generalise the inequality in Eq.~(\ref{eq:vectorial_result}) to all non-entangled states by considering a mixture of separable states i.e. $\hat{\rho}^{AB} = \sum_k P_k \hat{\rho}^{AB}_k$ where $P_k$ is the probability of $\hat{\rho}^{AB}_k = \hat{\rho}^A_k \otimes \hat{\rho}^B_k$. The singlet fraction $p_{sk}$ of each $\hat{\rho}^{AB}_k$ is still bounded by Eq.~(\ref{eq:vectorial_result}), thus
\begin{equation}
\label{eq:general_vectorial_result}
\overline{p_s} = \sum_k P_kp_{sk} \leq \frac{1-\sum_k P_k m_k^2}{2} = \frac{1-\overline{m^2}}{2}
\end{equation}
since $\overline{m^2} = \sum_k P_k m_k^2$. Hence if the two-body state is non-entangled i.e.~$\hat{\rho}^{AB} = \sum_k P_k \hat{\rho}^A_k \otimes \hat{\rho}^B_k$, then $p_s \leq (1-m^2)/2$ holds.
 
\section{Concurrence of entangled states \label{sec:proof}}

The contrapositive must also be true: if $p_s > (1-m^2)/2$, then $\hat{\rho}^{AB}$ is entangled. In fact, we find the concurrence \cite{wootters_entanglement_1998} of $\hat{\rho}^{AB}$ can be bounded using $p_s$ and $m$, without any additional assumptions.

First, we define a ``spun state'' as the state $\hat{\rho}^{AB}$ averaged uniformly over local rotations about the $z$ axis, $U_z(\theta) = U_z^A(\theta) \otimes U_z^B(\theta)$:
\begin{equation}
\label{eq:spinning}
\braket{\hat{\rho}^{AB}} = \frac{1}{2\pi} \int_0^{2\pi} d\theta \, \hat{U}^\dagger_z(\theta) \hat{\rho}^{AB} \hat{U}_z(\theta).
\end{equation}
This transformation eliminates coherences between states in $\hat{\rho}^{AB}$ with different angular momentum quantum number $S_z$, since $\hat{U}_z(\theta) = \exp{[i\theta {\hat{S}}_z]}$. Populations and coherence between $\ket{s_0}$ and $\ket{t_0}$ are unaffected, leaving 
\begin{equation}
\label{eq:spun_matrix}
\braket{\hat{\rho}^{AB}} = p_s \ket{s_0}\bra{s_0} + q_0\ket{s_0}\bra{t_0} + q_0^*\ket{t_0}\bra{s_0} + \sum_{i \in \left \{ 0, \pm 1 \right \}} (p_{ii}\ket{t_i}\bra{t_i})
\end{equation}
which now has only six degrees of freedom. Crucially, because rotation can be implemented using local operation and classical communication (LOCC), the spun state is at most as entangled as the unspun state i.e.~$\mathcal{C}(\hat{\rho}^{AB}) \geq \mathcal{C}(\braket{\hat{\rho}^{AB}})$ \cite{nielsen_quantum_2000}. 

\pagebreak 

Next, we constrain the state to have polarisation $m$. Choosing the $z$ axis along the measured direction of $\bvec{m}$,
\begin{eqnarray}
\label{eq:spun_matrix2}
\braket{\hat{\rho}^{AB}} = & p_s\ket{s_0}\bra{s_0} + a\ket{t_0}\bra{t_0} + ce^{i\phi}\ket{s_0}\bra{t_0} \nonumber \\
&+ ce^{-i\phi}\ket{t_0}\bra{s_0} + \frac{b+m}{2}\ket{t_1}\bra{t_1} + \frac{b-m}{2}\ket{t_{-1}}\bra{t_{-1}},
\end{eqnarray}
where the normalised populations are $p_s + a + b = 1$ and the coherence is $c=\eta\sqrt{ap_s}$ with $\eta \in [0,1]$. 

\begin{figure}[t!] \begin{center}
\includegraphics[width=0.5\textwidth]{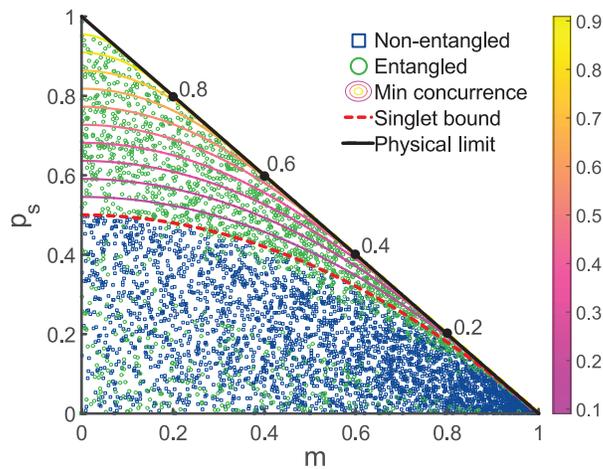} 
\caption{Each circle in singlet fraction $p_s$ vs. magnetisation $m$ space is a randomly generated state $\braket{\hat{\rho}^{AB}}$ described by Eq.~(\ref{eq:spun_matrix}). Blue squares have $\mathcal{C} = 0$, while green circles have $\mathcal{C}>0$. Note that there are blue squares immediately beneath the singlet bound Eq.~(\ref{eq:singlet_bound}) while there are none above, evidence that the bound is a tight and sufficient condition for entanglement. About the singlet bound, contour lines of $\mathcal{C}$ give the minimum concurrence (indicated by the colour scale) of a state with a given $p_s$ and $m$. For a given $\mathcal{C}$, the line of minimum $p_s$ follows Eq.~(\ref{eq:conc_contour}) from 
$p_s = (1+ \mathcal{C})/2$ at $m=0$ to $p_s = \mathcal{C}$ at $m=1 - \mathcal{C}$. Several $\mathcal{C}$ values are given along physical limit (Eq.~\ref{eq:physical_bound}).}
\label{fig:random_states}
\end{center} \end{figure}

Finally, we explicitly compute the concurrence of $\braket{\hat\rho^{AB}}$. The eigenvalues of the matrix $R = \left[{\sqrt{\braket{\hat\rho^{AB}}}\braket{\widetilde{\hat\rho^{AB}}}\sqrt{\braket{\hat\rho^{AB}}}}\right]^{1/2}$, where $\braket{\widetilde{\hat\rho^{AB}}}=(\sigma_y\otimes\sigma_y)\braket{\hat\rho^{AB}}^*(\sigma_y\otimes\sigma_y)$ is the ``spin-flipped'' state, are
\begin{eqnarray}
\fl \qquad \lambda_{1,2} =  \frac{1}{\sqrt{2}}\Big[a^2 + p_s^2 - 2c^2\cos{2\phi}
\pm \sqrt{(a^2 + p_s^2 - 2c^2\cos{2\phi})^2 - 4(c^2 - ap_s)^2}\Big]^{1/2}, \\
\fl \qquad \lambda_{3} = \lambda_{4} = \case{1}{2} \sqrt{b^2 - m^2}. 
\end{eqnarray}
The concurrence is then
\begin{eqnarray} 
\mathcal{C}(\braket{\hat{\rho}^{AB}}) &= \max[0, \, \lambda_1 - \lambda_2 - \lambda_3 - \lambda_4] \nonumber \\
&= \max[0, \, \sqrt{(p_s - a)^2 + 4c^2\sin^2{\phi}} - \sqrt{b^2 - m^2}], \label{eq:general_expression_conc}
\end{eqnarray}
which is nonzero when 
\begin{equation}
\label{eq:explicit_conc}
p_s > \frac{1}{2} \left( \frac{1-2a-m^2}{1-2a+2a\eta^2\sin^2{\phi}} \right).
\end{equation}
Since $\mathcal{C}(\hat{\rho}^{AB}) \geq \mathcal{C}(\braket{\hat{\rho}^{AB}})$, Eq.~(\ref{eq:explicit_conc}) provides a general bound on the singlet fraction for the entanglement of any $\hat{\rho}^{AB}$ with magnetisation $m$, triplet population $a$, and coherence $c$. With only the observables $p_s$ and $m$, this yields a sufficient condition for entanglement:
\begin{equation}
\label{eq:singlet_bound_conc}
p_s > \sup_{a, \eta, \phi}\left[\frac{1}{2}\frac{1-2a-m^2}{1-2a+2a\eta\sin^2{\phi}}\right]= \frac{1-m^2}{2}.
\end{equation}
With $a=0$ (which implies $c=0$) and $p_s = (1-m^2)/2$, we see that Eq.~(\ref{eq:spun_matrix2}) reduces to separable pure states, which fulfils Eq.~(\ref{eq:vectorial_result}) and saturates the bound. Another special case is the Werner state $a=b/2=(1-p_s)/3$, $m=0$, and $c=0$, for which Eq.~(\ref{eq:explicit_conc}) becomes  $p_s>1/2$.

The singlet bound found here (Eq.~\ref{eq:singlet_bound}) improves upon the generalised witness of Ref.~\cite{mintert_observable_2007}, which when applied to $\braket{\hat{\rho}^{AB}}$ with $a=0$, yields the sufficient condition $p_s \geq (1+\sqrt{1 + 3\mathcal{C}^2})/3$. 

The bound can be generalised to a threshold for finite concurrence, knowing only $m$ and $p_s$, by noting that the minimum of Eq.~(\ref{eq:general_expression_conc}) occurs when $a=0$. Along with the constraint of a normalised probability, $p_s+b=1$, this gives Eq.~(\ref{eq:minconc}). 
Solving for $p_s$, this gives a tight and sufficient condition for $\hat{\rho}^{AB}$ having at least concurrence $\mathcal{C}$, namely
\begin{equation}
\label{eq:conc_contour}
p_s \geq \frac{1-\mathcal{C}^2 - m^2}{2(1-\mathcal{C})} 
\qquad  \mbox{and}  \qquad
m \leq 1 - \mathcal{C}
\end{equation}
where Eq.~(\ref{eq:singlet_bound}) is now found from the condition $\mathcal{C}>0$. Equations (\ref{eq:minconc}) and (\ref{eq:conc_contour}) are the central results of our work.

We verify these relations by generating random mixed states that span the $p_s$ and $m$ space, and computing their concurrence. Each point in Fig.~\ref{fig:random_states} corresponds to one of five thousand random spun mixed states. The blue squares have $\mathcal{C}(\braket{\hat{\rho}^{AB}}) = 0$ and are not entangled while green circles have $\mathcal{C}(\braket{\hat{\rho}^{AB}})>0$ and are entangled. All points lie within the physical limit, Eq.~(\ref{eq:physical_bound}). The absence of non-entangled states above the singlet bound demonstrates that Eq.~(\ref{eq:singlet_bound}) is a sufficient condition for entanglement of $\hat{\rho}^{AB}$, while the existence of non-entangled states immediately beneath the bound demonstrates the tightness of the condition. Note that there are also entangled states below the singlet bound, as it is not a necessary condition for entanglement. Figure~\ref{fig:random_states} also shows contour lines of minimum $\mathcal{C}$ determined from a larger set of random matrices. The locus of points with at least concurrence $\mathcal{C}$ or greater is bounded by Eqs.~(\ref{eq:conc_contour}) and (\ref{eq:physical_bound}). 

\section{Discussion \label{sec:discussion}}

Several recent experimental works can be re-interpreted in light of our results. We will consider three measurements sensitive to $p_s$: mapping onto vibrational states in a superlattice, sweep-projection onto singlet dimers, and measuring the s-wave contact. We focus on experiments with fermions, even though our results apply to mixtures with any exchange statistics.

Controlled collisions in optical superlattices have been used both to create and to detect pairwise entanglement 
\cite{Rey:2007gc,anderlini_controlled_2007,Trotzky:2008jy,Trotzky:2010ka,Greif:2011fr,Nascimbene:2012fp,greif_short-range_2013}. However, when the effect of uncontrolled collisions are measured with the same technique, the efficiency of observing $p_s$ may be hampered by a randomised choice of pairs, if a simple lattice is pairwise projected into the superlattice. For instance, Greif {\em et al.} \cite{greif_short-range_2013} find that in a dimerised lattice, the singlet fraction of fermion pairs is at least $p_s = 0.31$. This was an effective probe of spin correlations, but insufficient to prove entanglement by Eq.~(\ref{eq:singlet_bound}).

The association of atomic fermions into s-wave pairs can also requires an initially singlet spin state. Thus efficiency $\mathcal{P}$ of association is a lower bound on $p_s$. For example, sweeping the magnetic field across a Feshbach resonance in experiments with unpolarised Fermi gases of $^{40}$K and $^6$Li, $\mathcal{P}$ as high as 85\% is observed \cite{Cubizolles:2003fn,Greiner:2003bj,hodby_production_2005}. This surpasses the 50\% upper limit discussed in Refs.~\cite{pazy_conversion_2004,Chwedenczuk:2004bm} which is also seen as an apparent limit in some experiments \cite{strecker_conversion_2003,Regal:2003ex}. We interpret this limit as $p_s=0.5$, which is the maximum singlet fraction of a non-entangled state: experiments (and theoretical treatments) finding $\mathcal{P} \leq 0.5$ use separable states, whereas experiments observing $\mathcal{P} > 0.5$ allow multiple collisions to occur before or during the magnetic field ramp. In some conditions, these collisions have produced pairwise entanglement. From Eq.~(\ref{eq:minconc}), we can infer that the concurrence was $\mathcal{C}\geq0.7$ for $p_s \geq \mathcal{P} \approx 0.85$ in Refs.~\cite{Greiner:2003bj,Cubizolles:2003fn,hodby_production_2005}.

Pairwise-entangled states of an unpolarised Fermi gas are not surprising: in a weakly interacting Fermi s-wave superfluid, each spin-up fermion is (monogamously) entangled with a spin-down partner. However entanglement dynamics in a polarised gas is an active topic of discussion. Calculations of the s-wave contact $\mathcal{I}$ in a polarised Fermi gas  \cite{bardon_transverse_2014,He:2016vm} have shown that $\mathcal{I} \propto 1-m^2$ at high temperature, and $\mathcal{I} \propto 1-m$ at low temperature. Since $\mathcal{I}$ reflects interaction strength, which in turn requires spin-singlet wave functions between fermions, this is similar to a study of $p_s$ versus $m$. The conversion of $\mathcal{I}$ to an absolute value of $p_s$ requires many-body theory and precise knowledge of density, temperature, and interaction strength. For this reason the spin correlations found by Bardon {\em et al.} \cite{bardon_transverse_2014} using $\mathcal{I}$ and $m$, for instance, cannot easily be classified using the singlet bound. More clear would be to study demagnetisation dynamics using association efficiency $\mathcal{P}$. One would anticipate a temperature threshold, below which the gas evolves from a separable state to a pairwise-entangled state through random collisions.

In sum, we have established a sufficient and tight condition for bipartite entanglement between spin degrees of freedom in an arbitrary system of spins, without any assumption of equilibrium, population balance, or symmetry. We find that the concurrence can be bounded simply by the magnetisation and singlet fraction, through Eq.~(\ref{eq:minconc}). This enables the distinction between classical spin correlations and necessarily quantum correlations in ensembles of ultracold fermions.

\ack
The authors thank D.\ DeMille, A.\ Dua, Xiwen Guan, B.\ Sanders, C.\ Simon, A.\ M.\ Steinberg, E.\ Taylor, S.\ Trotzky, Shizhong Zhang, and Huaixiu Zheng for stimulating conversations, as well as K.\ Heshami for assistance with the manuscript. L.\ J.\ acknowledges support from ARO, AFOSR MURI, the Alfred P. Sloan Foundation, and the Packard Foundation. J.\ T.\ acknowledges support from AFOSR, ARO, and NSERC.

\section*{\bf References}

\bibliography{refsSingletBound}

\end{document}